\documentclass[prl,twocolumn,twoside,preprintnumbers,superscriptaddress,nofootinbib,showpacs,showkeys]{revtex4}
\usepackage{amsmath,slashed}
\usepackage{graphicx,graphics}
\usepackage{dcolumn}
\usepackage{bm}
\usepackage[hyperfootnotes=false]{hyperref}
\usepackage{xspace}

\newcommand{\Fpi}{F_\pi}
\newcommand{\mpi}{M_{\pi}}

\newcommand{\ga}{g_A}
\newcommand{\Order}{\mathcal{O}}

\newcommand{\mN}{m_N}

\newcommand{\GeV}{\,\text{GeV}}
\newcommand{\beq}{\begin{equation}}
\newcommand{\eeq}{\end{equation}}

\allowdisplaybreaks[1]

\begin{document}

\preprint{INT-PUB-15-034}
\title{Matching pion--nucleon Roy--Steiner equations to chiral perturbation theory}

\author{Martin Hoferichter}
\affiliation{Institut f\"ur Kernphysik, Technische Universit\"at Darmstadt, D--64289 Darmstadt, Germany}
\affiliation{ExtreMe Matter Institute EMMI, GSI Helmholtzzentrum f\"ur Schwerionenforschung GmbH, D--64291 Darmstadt, Germany}
\affiliation{Institute for Nuclear Theory, University of Washington, Seattle, WA 98195-1550, USA}
\author{Jacobo Ruiz de Elvira}
\author{Bastian Kubis}
\affiliation{Helmholtz--Institut f\"ur Strahlen- und Kernphysik (Theorie) and\\
   Bethe Center for Theoretical Physics, Universit\"at Bonn, D--53115 Bonn, Germany}
\author{Ulf-G.\ Mei{\ss}ner}
\affiliation{Helmholtz--Institut f\"ur Strahlen- und Kernphysik (Theorie) and\\
   Bethe Center for Theoretical Physics, Universit\"at Bonn, D--53115 Bonn, Germany}
\affiliation{Institut f\"ur Kernphysik, Institute for Advanced Simulation, 
   J\"ulich Center for Hadron Physics, JARA-HPC, and JARA-FAME,  Forschungszentrum J\"ulich, D--52425  J\"ulich, Germany}

\begin{abstract}
We match the results for the subthreshold parameters of pion--nucleon scattering obtained from a solution of Roy--Steiner equations
to chiral perturbation theory up to next-to-next-to-next-to-leading order, to extract the pertinent low-energy constants 
including a comprehensive analysis of systematic uncertainties and correlations.
We study the convergence of the chiral series by investigating the chiral expansion of threshold parameters up to the same order and discuss the role of the $\Delta(1232)$ resonance in this context.
Results for the low-energy constants are also presented in the counting scheme usually applied in chiral nuclear effective field theory,
where they serve as crucial input to determine the long-range part of the nucleon--nucleon potential as well as three-nucleon forces.
\end{abstract}


\maketitle

\section{Introduction}

Chiral symmetry of QCD, the invariance of the QCD Lagrangian under chiral 
rotations of the quark fields in the chiral limit of vanishing quark masses,
is a powerful tool to elucidate the properties of strong interactions at low 
energies, where QCD becomes non-perturbative. This chiral symmetry is known to be
broken spontaneously and explicitly, with the appearance of almost massless
pseudo-Goldstone bosons, the pions.
By expanding systematically around the chiral limit of vanishing quark/pion masses,
one obtains an expansion 
in momenta and quark masses, with non-analytic terms predicted and the 
effects of high-energy physics incorporated in low-energy constants (LECs). 
These LECs appear in different physical processes, so that once fixed in one 
process, they can be used to predict others. In particular, one can derive 
low-energy theorems that relate different observables, at a given order in the 
chiral expansion. This approach, Chiral Perturbation Theory (ChPT), was pioneered 
in the meson sector in~\cite{Weinberg:1978kz,Gasser:1983yg,Gasser:1984gg}, and 
manifold extensions have been worked out over the last decades. In particular, 
it has been extended to the single-baryon sector, 
see~\cite{Jenkins:1990jv,Bernard:1992qa,Bernard:2007zu}, with pion--nucleon 
($\pi N$) scattering as one of the most fundamental applications~\cite{Bernard:1992qa,Mojzis:1997tu,Fettes:1998ud,Buettiker:1999ap,Fettes:2000gb,Fettes:2000xg,Becher:2001hv,Gasparyan:2010xz,Alarcon:2012kn,Chen:2012nx}.

However, as first pointed out in~\cite{Weinberg:1990rz,Weinberg:1991um,Weinberg:1992yk}, constraints from 
chiral symmetry are by no means limited to systems 
with at most one nucleon: once so-called nucleon--nucleon ($NN$) reducible 
contributions are separated, the remaining irreducible parts of the $NN$ potential 
again permit a chiral expansion, despite the non-perturbative nature of the 
$NN$ interactions. In this way, Nuclear Chiral Effective Field Theory (ChEFT), 
the extension of ChPT to the multi-nucleon sector, has been developed as a 
powerful tool for a systematic, model-independent approach to nuclear forces, 
see~\cite{Epelbaum:2008ga,Machleidt:2011zz} for recent reviews. One particularly valuable feature 
of ChEFT concerns the prediction of a hierarchy between two- and multi-nucleon 
forces, with the $NN$ interactions starting at leading order (LO), three-nucleon 
forces are predicted to enter at next-to-next-to-leading (N$^2$LO) order, 
and even higher forces are accordingly suppressed~\cite{vanKolck:1994yi}.
Further,  the LECs appearing in the expansion relate different processes. 
In fact, the LECs that appear in $\pi N$ scattering determine the long-range 
part of the $NN$ potential  and the three-nucleon force. Accordingly, if 
sufficiently precise information on $\pi N$ scattering were available, 
the required input could be immediately used in multi-nucleon applications.
This is the aim of the present Letter. 

Such improved input for the $\pi N$ LECs is becoming increasingly urgent, 
since with higher orders in ChEFT being worked out the uncertainties in 
the $\pi N$ LECs are starting to significantly contribute to the error budget 
in some observables, see e.g.~\cite{Kruger:2013kua}.
In the past, several strategies have been pursued: extractions from $\pi N$ 
scattering data, either in terms of phase shifts~\cite{Fettes:2000xg,Krebs:2012yv} 
or cross sections~\cite{Wendt:2014lja}, determinations from $NN$ 
observables~\cite{Rentmeester:2003mf,Perez:2013jpa}, or a combination of 
both~\cite{Carlsson:2015vda}.
Moreover, in~\cite{Buettiker:1999ap} the matching with a reconstructed 
dispersive $\pi N$ amplitude was performed in the subthreshold region where 
ChPT is expected to converge best, but the extrapolation from the physical region 
still required input from $\pi N$ data (similarly, while starting from the 
subthreshold region, the LECs are determined from fits to phase shifts 
in~\cite{Gasparyan:2010xz}). Given that the long-range contributions are 
entirely determined by $\pi N$ physics, $\pi N$ scattering provides the cleanest 
access and offers, at least for most LECs, also the highest sensitivity for their 
extraction. However, such a program has been hampered by 
inconsistencies in the low-energy $\pi N$ data base, as exemplified by 
contradicting partial-wave analyses, 
the Karlsruhe--Helsinki~\cite{Koch:1980ay,Hoehler} and the 
GWU/SAID solutions~\cite{Workman:2012hx}.

In $\pi\pi$ scattering, a similar situation prevailed until the consequent 
use of Roy equations~\cite{Roy:1971tc}, a combination of constraints from 
analyticity, unitarity, and crossing symmetry in the form of coupled 
integral equations for the partial waves. This significantly advanced the 
knowledge of the low-energy $\pi\pi$ phase 
shifts~\cite{Ananthanarayan:2000ht,GarciaMartin:2011cn}. Indeed, the matching 
to ChPT then allowed for a very precise determination of the pertinent $\pi\pi$ 
LECs~\cite{Colangelo:2001df}. 
Meanwhile, Roy-equation techniques have been extended to other 
processes~\cite{Buettiker:2003pp,Hoferichter:2011wk}, in particular,
a similar program has been pursued for $\pi N$ scattering based on 
Roy--Steiner (RS) equations~\cite{Hite:1973pm,Ditsche:2012fv,Hoferichter:2012wf,Hoferichter:2015dsa}, making use of a high-accuracy extraction of the $\pi N$ scattering 
lengths from pionic atoms as an additional 
constraint~\cite{Gotta:2008zza,Strauch:2010vu,Hennebach:2014lsa,Baru:2010xn,Baru:2011bw}.
In this Letter, we work out the consequences of our RS solution for the $\pi N$ 
LECs by matching the RS and the ChPT representation of the $\pi N$ amplitude 
in the subthreshold region. The main advantages of such an approach are the following: 
first, the $\pi N$ amplitude in the subthreshold region is a polynomial in the 
Mandelstam variables (apart from the Born terms), so that the chiral series is 
expected to converge best there. In contrast to~\cite{Buettiker:1999ap}, we do 
not need additional input from the physical region, as in our case the 
subthreshold parameters follow from the RS solution alone.
Second, the matching amounts to equating the subthreshold parameters 
from~\cite{Hoferichter:2015dsa,Hoferichter:2015hva} with their chiral expansion, which reduces 
the determination of the LECs to an algebraic problem.  Third, we can use the 
comprehensive error analysis performed in~\cite{Hoferichter:2015dsa,Hoferichter:2015hva}, 
which translates to a full covariance matrix for the extracted LECs.

\section{Subthreshold parameters}

We start by specifying conventions for the process
\beq
\pi^a(q)+N(p)\to\pi^b(q')+N(p'),
\eeq
with pion isospin labels $a$, $b$ and Mandelstam variables
\beq
s=(p+q)^2,\qquad t=(p'-p)^2,\qquad u=(p-q')^2,
\eeq
fulfilling $s+t+u=2\mN^2+2\mpi^2$.
We parameterize the scattering amplitude as
\begin{align}
T^{ba}(\nu,t)&=\delta^{ba}T^+(\nu,t)+\frac{1}{2}[\tau^b,\tau^a]T^-(\nu,t),\notag\\
T^I(\nu,t)&=\bar{u}(p')\bigg\{D^I(\nu,t)-\frac{[\slashed q',\slashed q]}{4\mN}
B^I(\nu,t)\bigg\}u(p),
\end{align}
where  $\nu=(s-u)/(4\mN)$, the isospin index $I=\pm$ refers to isoscalar/isovector 
amplitudes, $\mN$ and $\mpi$ to the nucleon and pion mass, and $\tau^a$ denotes 
isospin Pauli matrices. Throughout, the amplitudes with a definite $I=\pm$ index 
are understood to be related to
the $\pi^\pm p\to\pi^\pm p$ charge channels according to
\beq
\label{isospin_def}
X^\pm\equiv \frac{1}{2}\big(X_{\pi^-p\to\pi^- p}\pm X_{\pi^+p\to\pi^+ p}\big),
\eeq
for $X\in\{D,B,\ldots\}$, and the nucleon and pion mass are identified with the 
masses of the proton and the charged pion, respectively, 
see~\cite{Hoferichter:2015dsa} and~\cite{Gasser:2002am,Hoferichter:2009ez,Hoferichter:2009gn,Hoferichter:2012bz} for a discussion of the pertinent isospin-breaking 
corrections.
As mentioned above, once the Born terms are subtracted, the amplitude in the 
subthreshold region becomes a polynomial in $\nu$ and $t$.
A particularly convenient representation is provided by the subthreshold expansion
\begin{align}
\bar D^\pm(\nu,t)&=\begin{pmatrix}1\\\nu\end{pmatrix}
\sum_{n,m=0}^\infty d_{mn}^\pm\nu^{2m}t^n,\notag\\
\bar B^\pm(\nu,t)&=\begin{pmatrix}\nu\\1\end{pmatrix}
\sum_{n,m=0}^\infty b_{mn}^\pm\nu^{2m}t^n,
\end{align}
where the upper/lower entry corresponds to $I=\pm$, 
and the 
Born-term-subtracted amplitudes are defined as
\beq
\bar X^\pm(\nu,t)=X^\pm(\nu,t)-X^\pm_\text{pv}(\nu,t),\quad
X\in\{D,B\},
\eeq
with
\begin{align}
B^\pm_\text{pv}(\nu,t)&=g^2\bigg(\frac{1}{\mN^2-s}\mp\frac{1}{\mN^2-u}\bigg)
-\frac{g^2}{2\mN^2}\begin{pmatrix}0\\1\end{pmatrix},\notag\\
D^\pm_\text{pv}(\nu,t)&=\frac{g^2}{\mN}\begin{pmatrix}1\\ 0\end{pmatrix}
+\nu B^\pm_\text{pv}(\nu,t),
\end{align}
where $g$ denotes the $\pi N$ coupling constant.

\begin{table}[t]
\renewcommand{\arraystretch}{1.3}
\centering
\begin{tabular}{cr|cr}\toprule
$d_{00}^+\,[\mpi^{-1}]$ & $-1.36(3)$ & $d_{00}^-\,[\mpi^{-2}]$ & $1.41(1)$\\
$d_{10}^+\,[\mpi^{-3}]$ & $1.16(2)$ & $d_{10}^-\,[\mpi^{-4}]$ & $-0.159(4)$\\
$d_{01}^+\,[\mpi^{-3}]$ & $1.16(2)$ & $d_{01}^-\,[\mpi^{-4}]$ & $-0.141(5)$\\
$d_{20}^+\,[\mpi^{-5}]$ & $0.196(3)$ & $b_{00}^-\,[\mpi^{-2}]$ & $10.49(11)$\\
$d_{11}^+\,[\mpi^{-5}]$ & $0.185(3)$ & $b_{10}^-\,[\mpi^{-4}]$ & $1.00(3)$\\
$d_{02}^+\,[\mpi^{-5}]$ & $0.0336(6)$ & $b_{01}^-\,[\mpi^{-4}]$ & $0.21(2)$\\
$b_{00}^+\,[\mpi^{-3}]$ & $-3.45(7)$ & &\\\botrule
\end{tabular}
\caption{Subthreshold parameters from the RS 
analysis~\cite{Hoferichter:2015dsa,Hoferichter:2015hva}.}
\label{tab:RS_subthr}
\end{table}

For the matching to ChPT at N$^3$LO (complete one-loop order) we need 
the $13$ subthreshold parameters listed in Table~\ref{tab:RS_subthr}.
The solution of the RS equations is obtained by minimizing a $\chi^2$-like 
function, defined as the difference between left- and right-hand side of the 
equations on a grid of points, with respect to the subtraction constants and 
the low-energy phase shifts. Most of the subthreshold parameters listed in 
Table~\ref{tab:RS_subthr} already appear as subtraction constants of the RS 
system, and thus follow as output from the RS solution,
while the remaining ones, $d_{20}^+$, $d_{11}^+$, and $d_{02}^+$, are calculated 
from sum rules afterwards.
The uncertainty estimates include a number of effects: first, the RS equations 
are valid only in a finite energy range below the so-called matching point and 
only a finite number of partial waves are included explicitly in the solution. 
We varied the input for the matching condition as well as for the energy region 
above the matching point and higher partial waves, both regarding different 
partial-wave analyses and truncations of the partial-wave expansion. Furthermore, 
we varied the input for the $\pi N$ coupling constant within 
$g^2/(4\pi)=13.7(2)$~\cite{Baru:2010xn,Baru:2011bw} and investigated the sensitivity 
to the parameterization of the low-energy phase shifts used in the solution. 
Second, we observed that the RS equations are more sensitive to some subthreshold 
parameters than others. To account for this effect, we generated a set of 
solutions corresponding to different starting values of the $\chi^2$-minimization, 
while imposing sum rules for the higher subthreshold parameters, and took the 
observed distribution as an additional source of uncertainty. Third, we propagated 
the errors in the scattering lengths, which crucially enter as constraints in the 
minimization, to the results for the subthreshold parameters. 
Taking everything together we obtain a $13\times 13$ covariance matrix that 
encodes uncertainties and correlations of the $13$ subthreshold parameters 
relevant for the matching to ChPT.

\section{Chiral expansion}

The chiral expansion for the subthreshold parameters is spelled out 
explicitly in~\cite{Becher:2001hv}, in particular
\begin{align}
\label{expansion_subthr}
 d_{00}^+&=-\frac{2 \mpi^2 (2 c_1-c_3)}{\Fpi^2}
 +\frac{\ga^2 \big(3+8 \ga^2\big) \mpi^3}{64 \pi  \Fpi^4}\\
 &+\mpi^4 \Bigg\{\frac{16 \bar e_{14}}{\Fpi^2}+\frac{3\ga^2 \big(1+6 \ga^2\big)}{64 \pi ^2 \Fpi^4 \mN}-\frac{2 c_1-c_3}{16 \pi ^2 \Fpi^4}\Bigg\},\notag\\
 d_{00}^-&=\frac{1}{2 \Fpi^2}
 +\frac{4 \mpi^2 (\bar d_1+\bar d_2+2 \bar d_5)}{\Fpi^2}+\frac{\ga^4 \mpi^2}{48 \pi ^2 \Fpi^4}\notag\\
 &-\mpi^3 \Bigg\{\frac{8+12 \ga^2+11 \ga^4}{128 \pi  \Fpi^4 \mN}-\frac{4 c_1+\ga^2 (c_3-c_4)}{4 \pi  \Fpi^4}\Bigg\},\notag
\end{align}
where $c_i$, $\bar d_i$, and $\bar e_i$ denote the NLO, N$^2$LO, N$^3$LO $\pi N$ 
LECs, respectively, $\Fpi$ the pion decay constant, and $\ga$ the axial coupling 
of the nucleon.
The conventions for the $\bar e_i$ correspond to the general 
classification~\cite{Fettes:2000gb} and the $c_i$ have been redefined to absorb 
a quark-mass renormalization, see~\cite{Fettes:2000xg,Krebs:2012yv}. Finally, 
the expressions in~\eqref{expansion_subthr} follow the standard counting in the 
single-nucleon sector, where the expansion parameter is given by $\Order(p)
=\{p,\mpi\}/\Lambda_\text{b}$, for momenta $p$ and the breakdown-scale 
$\Lambda_\text{b}\sim \Lambda_\chi\sim4\pi\Fpi\sim\mN\sim M_\rho\sim 1\GeV$.
In contrast, the breakdown-scale in few-nucleon applications is typically 
lower, $\Lambda_\text{b}\sim 0.6\GeV$, so that 
relativistic corrections are often counted as 
$\{p,\mpi\}/\mN=\Order(p^2)$~\cite{Weinberg:1991um}. 
As a consequence, in this counting one would drop the $1/\mN$ suppressed 
terms in~\eqref{expansion_subthr}. In this Letter, we consider both counting 
schemes, which we will refer to as standard and $NN$ counting in the following.
The full set of subthreshold parameters can be easily inverted for the LECs, 
for the result at different chiral orders see Table~\ref{tab:LECs} (masses, 
$\Fpi$, and $\ga$ are taken from~\cite{PDG}). The errors as propagated from 
the subthreshold parameters are tiny compared to the shifts observed between 
chiral orders: clearly, the dominant uncertainty now resides in the chiral 
expansion. For completeness, we also quote the N$^3$LO correlation 
coefficients, see Table~\ref{tab:corr}. Note that this table contains
the correlation matrices for the standard and the $NN$ counting and therefore
appears asymmetric.

\begin{table}[t]
\renewcommand{\arraystretch}{1.3}
\centering
\begin{tabular}{crrrr}\toprule
& NLO & N$^2$LO & N$^3$LO & N$^3$LO$^{NN}$\\
$c_1$ & $-0.74(2)$ & $-1.07(2)$ & $-1.11(3)$ & $-1.10(3)$\\
$c_2$ & $1.81(3)$ & $3.20(3)$ & $3.13(3)$ & $3.57(4)$\\
$c_3$ & $-3.61(5)$ & $-5.32(5)$ & $-5.61(6)$ & $-5.54(6)$\\
$c_4$ & $2.17(3)$ & $3.56(3)$ & $4.26(4)$ & $4.17(4)$\\
$\bar d_1+\bar d_2$ & --- & $1.04(6)$ & $7.42(8)$ & $6.18(8)$\\
$\bar d_3$ & --- & $-0.48(2)$ & $-10.46(10)$ & $-8.91(9)$\\
$\bar d_5$ & --- & $0.14(5)$ & $0.59(5)$ & $0.86(5)$\\
$\bar d_{14}-\bar d_{15}$ & --- & $-1.90(6)$ & $-13.02(12)$ & $-12.18(12)$\\
$\bar e_{14}$ & --- & --- & $0.89(4)$ & $1.18(4)$\\
$\bar e_{15}$ & --- & --- & $-0.97(6)$ & $-2.33(6)$\\
$\bar e_{16}$ & --- & --- & $-2.61(3)$ & $-0.23(3)$\\
$\bar e_{17}$ & --- & --- & $0.01(6)$ & $-0.18(6)$\\
$\bar e_{18}$ & --- & --- & $-4.20(5)$ & $-3.24(5)$\\
\botrule
\end{tabular}
\caption{Results for the $\pi N$ LECs at NLO, N$^2$LO, and N$^3$LO (standard and 
$NN$ counting only differ at N$^3$LO, except for NLO in $c_4$, which in the $NN$ scheme becomes $2.44(3)$). The results for the $c_i$, $\bar d_i$, 
and $\bar e_i$ are given in units of $\GeV^{-1}$, $\GeV^{-2}$, and $\GeV^{-3}$, 
respectively.}
\label{tab:LECs}
\end{table}

\begin{table*}[t]
\renewcommand{\arraystretch}{1.3}
\centering
\begin{tabular}{crrrr|rrrr|rrrrr}
\toprule
& $c_1$ & $c_2$ & $c_3$ & $c_4$ & $\bar d_1+\bar d_2$ & $\bar d_3$ & $\bar d_5$ & $\bar d_{14}-\bar d_{15}$ & $\bar e_{14}$ & $\bar e_{15}$ & $\bar e_{16}$ & $\bar e_{17}$ & $\bar e_{18}$ \\
$c_1$ & $1$ & $0.18$ & $0.58$ & $0.06$ & $-0.42$ & $0.71$ & $0.04$ & $0.47$ & $-0.59$ & $0.33$ & $-0.21$ & $-0.11$ & $-0.21$\\
$c_2$ & $-0.20$ & $1$ & $-0.64$ & $-0.01$ & $0.67$ & $-0.36$ & $-0.27$ & $-0.55$ & $0.56$ & $-0.59$ & $0.59$ & $0.21$ & $0.47$\\
$c_3$ & $0.58$ & $-0.86$ & $1$ & $0.04$ & $-0.86$ & $0.91$ & $0.16$ & $0.87$ & $-0.97$ & $0.68$ & $-0.60$ & $-0.24$ & $-0.46$\\
$c_4$ & $0.06$ & $-0.03$ & $0.04$ & $1$ & $0.18$ & $-0.22$ & $0.03$ & $-0.31$ & $-0.02$ & $0.07$ & $-0.08$ & $-0.61$ & $-0.63$\\\colrule
$\bar d_1+\bar d_2$ & $-0.42$ & $0.83$ & $-0.86$ & $0.18$ & $1$ & $-0.83$ & $-0.40$ & $-0.94$ & $0.88$ & $-0.77$ & $0.74$ & $0.23$ & $0.34$\\
$\bar d_3$ & $0.68$ & $-0.63$ & $0.90$ & $-0.25$ & $-0.83$ & $1$ & $0.05$ & $0.93$ & $-0.94$ & $0.53$ & $-0.47$ & $-0.07$ & $-0.17$\\
$\bar d_5$ & $0.04$ & $-0.28$ & $0.16$ & $0.03$ & $-0.40$ & $0.03$ & $1$ & $0.18$ & $-0.14$ & $0.40$ & $-0.29$ & $-0.18$ & $-0.29$\\
$\bar d_{14}-\bar d_{15}$ & $0.47$ & $-0.73$ & $0.87$ & $-0.31$ & $-0.94$ & $0.93$ & $0.18$& $1$ & $-0.91$ & $0.64$ & $-0.61$ & $-0.03$ & $-0.21$\\\colrule
$\bar e_{14}$ & $-0.60$ & $0.77$ & $-0.97$ & $-0.02$ & $0.88$ & $-0.94$ & $-0.13$ & $-0.91$ & $1$ & $-0.70$ & $0.65$ & $0.23$ & $0.43$\\
$\bar e_{15}$ & $0.33$ & $-0.72$ & $0.68$ & $0.07$ & $-0.77$ & $0.52$ & $0.40$ & $0.64$ & $-0.69$ & $1$ & $-0.97$ & $-0.28$ & $-0.65$\\
$\bar e_{16}$ & $-0.21$ & $0.67$ & $-0.60$ & $-0.08$ & $0.74$ & $-0.45$ & $-0.29$ & $-0.61$ & $0.65$ & $-0.97$ & $1$ & $0.29$ & $0.60$\\
$\bar e_{17}$ & $-0.11$ & $0.25$ & $-0.24$ & $-0.61$ & $0.23$ & $-0.05$ & $-0.18$ & $-0.03$ & $0.23$ & $-0.28$ & $0.29$ & $1$ & $0.19$\\
$\bar e_{18}$ & $-0.20$ & $0.55$ & $-0.46$ & $-0.63$ & $0.34$ & $-0.14$ & $-0.29$ & $-0.20$ & $0.42$ & $-0.65$ & $0.60$ & $0.19$& $1$\\
\botrule
\end{tabular}
\caption{Correlation coefficients at N$^3$LO in standard (upper-right triangle) 
and $NN$ (lower-left triangle) counting.}
\label{tab:corr}
\end{table*}

In general, the values for the LECs are expected to be $\Order(1)$, e.g.\ 
$c_i \sim \ga/\Lambda_\text{b}$~\cite{Bernard:2007zu} and 
significant departures indicate the presence of additional degrees of freedom. 
In the case of $c_{2-4}$ the main origin of their enhancement is well understood: 
while other resonances do contribute as well, it is primarily the presence 
of the $\Delta(1232)$ resonance that makes these LECs take unnaturally 
large values~\cite{Bernard:1995dp,Bernard:1996gq,Becher:1999he}.
Following~\cite{Becher:1999he}, we extract the $\Delta$ contributions to the 
individual subthreshold parameters from the corresponding tree-level $\Delta$-exchange 
diagrams and convert the result to the LECs. For the numerical analysis
we use the $\Delta$ coupling constant $g_{\pi N\Delta}=1.2$~\cite{Hoehler}, 
which lies right in the middle of the range $1.05$ (extracted from the 
$\Delta$ width~\cite{Hemmert:1997ye}) and $3\ga/(2\sqrt{2})=1.35$ (predicted by 
large $N_c$~\cite{Dashen:1993as}).
Keeping the full pion-mass dependence, we obtain the values shown in the first 
column of Table~\ref{tab:LECs_Delta}, while the second column
corresponds to the leading expansion in $\mpi$ and $m_\Delta-\mN$. Only in the 
latter case one recovers the relation 
$c_2^\Delta=-c_3^\Delta=2c_4^\Delta$~\cite{Bernard:1996gq}.

\begin{table}[t]
\renewcommand{\arraystretch}{1.3}
\centering
\begin{tabular}{crr|crr|crr}\toprule
$c_1^\Delta$ & $0.0$ & $0.0$ & $\bar d_1^\Delta+\bar d_2^\Delta$ & $1.9$ & $1.9$ & $\bar e_{14}^\Delta$ & $-0.4$ & $0.0$ \\
$c_2^\Delta$ & $1.6$ & $2.2$ & $\bar d_3^\Delta$ & $-0.9$ & $-1.9$ & $\bar e_{15}^\Delta$ & $-2.6$ & $-3.2$ \\
$c_3^\Delta$ & $-2.1$ & $-2.2$ & $\bar d_5^\Delta$ & $-0.4$ & $0.0$ & $\bar e_{16}^\Delta$ & $1.4$ & $3.2$\\
$c_4^\Delta$ & $1.2$ & $1.1$ & $\bar d_{14}^\Delta-\bar d_{15}^\Delta$ & $-2.9$ & $-3.7$& $\bar e_{17}^\Delta$ & $0.3$ & $0.0$ \\
 &&&&&& $\bar e_{18}^\Delta$ & $1.1$ & $1.6$ \\
\botrule
\end{tabular}
\caption{$\Delta$ contributions to the $\pi N$ LECs, in GeV units, for the 
full $\Delta$-exchange diagrams (first column) and to leading order in 
$\mpi$ and $m_\Delta-\mN$ (second column).}
\label{tab:LECs_Delta}
\end{table}

While the $\Delta$ can indeed explain a significant portion of the physical values 
of the $c_i$, its effect is too small to explain the large numbers for 
the $\bar d_i$  that appear at N$^3$LO (except for $\bar d_5$). The origin 
of this large shift can be traced back to the terms proportional to 
$\ga^2(c_3-c_4)\sim-16\GeV^{-1}$ in $d_{00}^-$ in~\eqref{expansion_subthr} 
(and similarly in $d_{10}^-$, $d_{01}^-$, and $b_{00}^+$). These terms mimic loop 
diagrams with $\Delta$ degrees of freedom. Our results show that if the $\Delta$ 
is not included explicitly, such contributions lead to a substantial 
renormalization of the LECs. Indeed, if we drop the $c_3-c_4$ loop terms, 
the $\bar d_i$ are reduced to $\bar d_{1+2,3,14-15}=(2.2,-3.9,-2.6)\GeV^{-2}$, in 
good agreement with the expectations from Table~\ref{tab:LECs_Delta}.

\section{Threshold parameters}

With the LECs determined by matching to the subthreshold expansion, it is important 
to check how well the chiral series converges in other kinematic regions. A prime 
test case is provided by the $S$-wave scattering lengths $a_{0+}^\pm$: they are known 
very precisely from pionic atoms~\cite{Baru:2010xn,Baru:2011bw}. In the isospin 
conventions~\eqref{isospin_def} their values are $a_{0+}^+=-0.9(1.4)$ and 
$a_{0+}^-=85.4(9)$ (always in units of $10^{-3}\mpi^{-1}$), and the problem is 
still purely algebraic.  The fourth-order expressions for their chiral expansion 
were first given in~\cite{Bernard:1995pa}, in our conventions they read
\begin{align}
\label{aminus}
 a_{0+}^+&=\frac{\mpi^2}{4\pi  \Fpi^2 (\mN+\mpi)}\bigg\{
 \frac{3 \ga^2 \mN \mpi}{64 \pi\Fpi^2}-\frac{\ga^2 \mpi^2}{16\mN^2}\notag\\
 &-\frac{1}{4}\big[\ga^2+8 \mN (2 c_1-c_2-c_3)\big]\notag\\
 &+\mpi^2 \big[-16 c_1 c_2+\bar d_{18} \ga+16 \mN (\bar e_{14}+\bar e_{15}
+\bar e_{16})\big]\notag\\
 &-\frac{\mpi^2 \big[8-3\ga^2+2 \ga^4+4 \mN (2 c_1-c_3)\big]}
{64 \pi ^2 \Fpi^2}\bigg\},\notag\\
 a_{0+}^-&=\frac{\mN \mpi}{8 \pi  \Fpi^2 (\mN+\mpi)}\bigg\{1
+\frac{\mpi^2}{8\pi^2\Fpi^2} +\frac{\ga^2 \mpi^2}{4\mN^2}\notag\\
 &+8\mpi^2 (\bar d_1+\bar d_2+\bar d_3+2 \bar d_5)\bigg\}.
\end{align}
Fixing the only new LEC, $\bar d_{18}$, from the Goldberger--Treiman discrepancy,
\beq
\bar d_{18}=\frac{\ga}{2\mpi^2}\bigg(1-\frac{g\Fpi}{\mN\ga}\bigg)
=-0.44(24)\GeV^{-2},
\eeq
we obtain the following results
\begin{align}
 a_{0+}^+&=\{-23.8,0.2,-7.9\},\{-14.2,0.2,-1.4\},\notag\\
 a_{0+}^-&=\{79.4,92.9,59.4\},\{79.4,92.2,69.2\},
\end{align}
where the first/second array refers to the standard/$NN$ counting and the 
three entries to NLO, N$^2$LO, N$^3$LO.
It is not surprising that the chiral expansion in the isoscalar sector is 
slow, after all its LO vanishes. Unexpectedly, a similarly slow convergence
is  also found for $a_{0+}^-$, whose low-energy theorem at LO is tantalizingly 
close to the full answer, while the agreement in both counting schemes deteriorates 
when going to fourth order. The largest part of this discrepancy can be attributed 
to the $\Delta$ loops discussed above, i.e.\ for $a_{0+}^-$ the largest portion 
of the $c_3-c_4$ terms does cancel between $\bar d_1+\bar d_2$ and $\bar d_3$, but 
the cancellation is incomplete and the remainder spoils the agreement with the 
pionic-atom value.

This example shows that in a theory without explicit $\Delta$ degrees of freedom 
the LECs determined in a particular kinematic region do not necessarily ensure 
convergence in the full low-energy domain. However, especially when going to
higher orders, including the $\Delta$ explicitly
becomes extremely challenging, 
so that in practice the $\Delta$-less approach can be pushed to higher orders  
and it remains to be seen if in the end the $\Delta$-full or $\Delta$-less
theory proves more efficient.
We argue here that for $\Delta$-less applications in $NN$ scattering 
matching at the subthreshold point 
is the preferred choice: the two-pion-exchange diagrams can be reconstructed 
from $\pi N$ scattering by means of Cutkosky rules~\cite{Kaiser:2001pc} 
(see~\cite{Entem:2014msa,Epelbaum:2014sza} for recent applications of this 
approach), with spectral functions involving $\pi N$ amplitudes either directly 
evaluated at or weighted towards zero pion center-of-mass 
momenta~\cite{Kaiser:2001pc}, which translates to $s=\mN^2-\mpi^2$. Moreover, 
for physical values of the momentum transfer $t$ in $NN$ scattering the Cauchy 
kernels in the spectral integrals become largest for $t=0$. Since the corresponding 
combination of $(s,t)$ is much closer to subthreshold $(\mN^2+\mpi^2,0)$ than 
threshold $((\mN+\mpi)^2,0)$ kinematics, we conclude that the LECs to be applied 
in nuclear forces should be extracted from the subthreshold point in $\pi N$ 
scattering instead of the physical region. In the present Letter we have presented 
such an extraction based on a comprehensive analysis of low-energy $\pi N$ 
scattering in the framework of RS equations.
The corresponding LECs clearly defined at a given chiral order 
will be valuable for assessing the uncertainties from the long-range part of the nuclear force 
in future ChEFT calculations~\cite{Epelbaum:2014efa,Furnstahl:2015rha}.

\section*{Acknowledgments}

We thank V.~Bernard, E.~Epelbaum, A.~Gasparyan, H.~Krebs, A.~Schwenk, and D.~Siemens for helpful discussions
and comments on the manuscript.
Financial support by
BMBF ARCHES, the Helmholtz Virtual Institute NAVI (VH-VI-417),
the DFG (SFB/TR 16, ``Subnuclear Structure of Matter''),
and  the DOE (Grant No.\ DE-FG02-00ER41132) 
is gratefully acknowledged.


\begin{thebibliography}{99}

\bibitem{Weinberg:1978kz} 
  S.~Weinberg,
  Physica A {\bf 96}, 327 (1979).

\bibitem{Gasser:1983yg} 
  J.~Gasser and H.~Leutwyler,
  Annals Phys.\  {\bf 158}, 142 (1984).
  
\bibitem{Gasser:1984gg} 
  J.~Gasser and H.~Leutwyler,
  Nucl.\ Phys.\ B {\bf 250}, 465 (1985).
  
\bibitem{Jenkins:1990jv} 
  E.~E.~Jenkins and A.~V.~Manohar,
  Phys.\ Lett.\ B {\bf 255}, 558 (1991).
  
\bibitem{Bernard:1992qa} 
  V.~Bernard, N.~Kaiser, J.~Kambor and U.-G.~Mei\ss ner,
  Nucl.\ Phys.\ B {\bf 388}, 315 (1992).
  
\bibitem{Bernard:2007zu} 
  V.~Bernard,
  Prog.\ Part.\ Nucl.\ Phys.\  {\bf 60}, 82 (2008)
  [arXiv:0706.0312 [hep-ph]].
 
\bibitem{Mojzis:1997tu} 
  M.~Moj\v zi\v s,
  Eur.\ Phys.\ J.\ C {\bf 2}, 181 (1998)
  [hep-ph/9704415].
 
\bibitem{Fettes:1998ud} 
  N.~Fettes, U.-G.~Mei\ss ner and S.~Steininger,
  Nucl.\ Phys.\ A {\bf 640}, 199 (1998)
  [hep-ph/9803266].
  
\bibitem{Buettiker:1999ap} 
  P.~B\"uttiker and U.-G.~Mei\ss ner,
  Nucl.\ Phys.\ A {\bf 668}, 97 (2000)
  [hep-ph/9908247].
 
\bibitem{Fettes:2000gb} 
  N.~Fettes, U.-G.~Mei\ss ner, M.~Moj\v zi\v s and S.~Steininger,
  Annals Phys.\  {\bf 283}, 273 (2000)
  [Annals Phys.\  {\bf 288}, 249 (2001)]
  [hep-ph/0001308].
 
\bibitem{Fettes:2000xg} 
  N.~Fettes and U.-G.~Mei\ss ner,
  Nucl.\ Phys.\ A {\bf 676}, 311 (2000)
  [hep-ph/0002162].
  
\bibitem{Becher:2001hv} 
  T.~Becher and H.~Leutwyler,
  JHEP {\bf 0106}, 017 (2001)
  [hep-ph/0103263].
  
\bibitem{Gasparyan:2010xz} 
  A.~Gasparyan and M.~F.~M.~Lutz,
  Nucl.\ Phys.\ A {\bf 848}, 126 (2010)
  [arXiv:1003.3426 [hep-ph]].
  
\bibitem{Alarcon:2012kn} 
  J.~M.~Alarc\'on, J.~Martin Camalich and J.~A.~Oller,
  Annals Phys.\  {\bf 336}, 413 (2013)
  [arXiv:1210.4450 [hep-ph]].
  
\bibitem{Chen:2012nx} 
  Y.~H.~Chen, D.~L.~Yao and H.~Q.~Zheng,
  Phys.\ Rev.\ D {\bf 87}, 054019 (2013)
  [arXiv:1212.1893 [hep-ph]].

\bibitem{Weinberg:1990rz} 
  S.~Weinberg,
  Phys.\ Lett.\ B {\bf 251}, 288 (1990).
  
\bibitem{Weinberg:1991um} 
  S.~Weinberg,
  Nucl.\ Phys.\ B {\bf 363}, 3 (1991).
  
\bibitem{Weinberg:1992yk} 
  S.~Weinberg,
  Phys.\ Lett.\ B {\bf 295}, 114 (1992)
  [hep-ph/9209257].
  
\bibitem{Epelbaum:2008ga} 
  E.~Epelbaum, H.~W.~Hammer and U.-G.~Mei\ss ner,
  Rev.\ Mod.\ Phys.\  {\bf 81}, 1773 (2009)
  [arXiv:0811.1338 [nucl-th]].
  
\bibitem{Machleidt:2011zz}
  R.~Machleidt and D.~R.~Entem,
  Phys.\ Rept.\  {\bf 503} (2011) 1
  [arXiv:1105.2919 [nucl-th]].

\bibitem{vanKolck:1994yi} 
  U.~van Kolck,
  Phys.\ Rev.\ C {\bf 49}, 2932 (1994).
  
\bibitem{Kruger:2013kua} 
  T.~Kr\"uger, I.~Tews, K.~Hebeler and A.~Schwenk,
  Phys.\ Rev.\ C {\bf 88}, 025802 (2013)
  [arXiv:1304.2212 [nucl-th]].
 
\bibitem{Krebs:2012yv} 
  H.~Krebs, A.~Gasparyan and E.~Epelbaum,
  Phys.\ Rev.\ C {\bf 85}, 054006 (2012)
  [arXiv:1203.0067 [nucl-th]].
 
\bibitem{Wendt:2014lja} 
  K.~A.~Wendt, B.~D.~Carlsson and A.~Ekstr\"om,
  arXiv:1410.0646 [nucl-th].
  
\bibitem{Rentmeester:2003mf} 
  M.~C.~M.~Rentmeester, R.~G.~E.~Timmermans and J.~J.~de Swart,
  Phys.\ Rev.\ C {\bf 67}, 044001 (2003)
  [nucl-th/0302080].
  
\bibitem{Perez:2013jpa} 
  R.~Navarro P\'erez, J.~E.~Amaro and E.~Ruiz Arriola,
  Phys.\ Rev.\ C {\bf 88}, 064002 (2013)
  [Phys.\ Rev.\ C {\bf 91}, 029901 (2015)]
  [arXiv:1310.2536 [nucl-th]].
  
\bibitem{Carlsson:2015vda} 
  B.~D.~Carlsson {\it et al.},
  arXiv:1506.02466 [nucl-th].
 
\bibitem{Koch:1980ay} 
  R.~Koch and E.~Pietarinen,
  Nucl.\ Phys.\ A {\bf 336}, 331 (1980).
 
\bibitem{Hoehler}
  G.~H\"{o}hler,
  {\it Pion--Nukleon-Streuung: Methoden und Ergebnisse},
  in Landolt-B\"ornstein, {\bf 9b2}, ed.\ H.~Schopper,
  Springer Verlag, Berlin, 1983. 
  
\bibitem{Workman:2012hx} 
  R.~L.~Workman, R.~A.~Arndt, W.~J.~Briscoe, M.~W.~Paris and I.~I.~Strakovsky,
  Phys.\ Rev.\ C {\bf 86}, 035202 (2012)
  [arXiv:1204.2277 [hep-ph]].
  
\bibitem{Roy:1971tc} 
  S.~M.~Roy,
  Phys.\ Lett.\ B {\bf 36}, 353 (1971).
  
\bibitem{Ananthanarayan:2000ht} 
  B.~Ananthanarayan, G.~Colangelo, J.~Gasser and H.~Leutwyler,
  Phys.\ Rept.\  {\bf 353}, 207 (2001)
  [hep-ph/0005297].
  
\bibitem{GarciaMartin:2011cn} 
  R.~Garc\'ia-Mart\'in, R.~Kami\'nski, J.~R.~Pel\'aez, J.~Ruiz de Elvira and F.~J.~Yndur\'ain,
  Phys.\ Rev.\ D {\bf 83}, 074004 (2011)
  [arXiv:1102.2183 [hep-ph]].

\bibitem{Colangelo:2001df} 
  G.~Colangelo, J.~Gasser and H.~Leutwyler,
  Nucl.\ Phys.\ B {\bf 603}, 125 (2001)
  [hep-ph/0103088].
  
\bibitem{Buettiker:2003pp} 
  P.~B\"uttiker, S.~Descotes-Genon and B.~Moussallam,
  Eur.\ Phys.\ J.\ C {\bf 33}, 409 (2004)
  [hep-ph/0310283].
  
\bibitem{Hoferichter:2011wk} 
  M.~Hoferichter, D.~R.~Phillips and C.~Schat,
  Eur.\ Phys.\ J.\ C {\bf 71}, 1743 (2011)
  [arXiv:1106.4147 [hep-ph]].
  
\bibitem{Hite:1973pm} 
  G.~E.~Hite and F.~Steiner,
  Nuovo Cim.\ A {\bf 18}, 237 (1973).
  
\bibitem{Ditsche:2012fv} 
  C.~Ditsche, M.~Hoferichter, B.~Kubis and U.-G.~Mei{\ss}ner,
  JHEP {\bf 1206}, 043 (2012)
  [arXiv:1203.4758 [hep-ph]].
  
\bibitem{Hoferichter:2012wf} 
  M.~Hoferichter, C.~Ditsche, B.~Kubis and U.-G.~Mei{\ss}ner,
  JHEP {\bf 1206}, 063 (2012)
  [arXiv:1204.6251 [hep-ph]].
  
\bibitem{Hoferichter:2015dsa} 
  M.~Hoferichter, J.~Ruiz de Elvira, B.~Kubis and U.-G.~Mei{\ss}ner,
  Phys.\ Rev.\ Lett.\  {\bf 115}, 092301 (2015)
  [arXiv:1506.04142 [hep-ph]].
  
\bibitem{Gotta:2008zza} 
  D.~Gotta {\it et al.},
  Lect.\ Notes Phys.\  {\bf 745}, 165 (2008).
  
\bibitem{Strauch:2010vu} 
  T.~Strauch {\it et al.},
  Eur.\ Phys.\ J.\ A {\bf 47}, 88 (2011)
  [arXiv:1011.2415 [nucl-ex]].
  
\bibitem{Hennebach:2014lsa} 
  M.~Hennebach {\it et al.},
  Eur.\ Phys.\ J.\ A {\bf 50}, 190 (2014)
  [arXiv:1406.6525 [nucl-ex]].
  
\bibitem{Baru:2010xn} 
  V.~Baru {\it et al.},
  Phys.\ Lett.\ B {\bf 694}, 473 (2011)
  [arXiv:1003.4444 [nucl-th]].
  
\bibitem{Baru:2011bw} 
  V.~Baru {\it et al.},
  Nucl.\ Phys.\ A {\bf 872}, 69 (2011)
  [arXiv:1107.5509 [nucl-th]].
  
\bibitem{Hoferichter:2015hva}
  M.~Hoferichter, J.~Ruiz de Elvira, B.~Kubis and U.-G.~Mei{\ss}ner,
  arXiv:1510.06039 [hep-ph].
 
\bibitem{Gasser:2002am} 
  J.~Gasser, M.~A.~Ivanov, E.~Lipartia, M.~Moj\v zi\v s and A.~Rusetsky,
  Eur.\ Phys.\ J.\ C {\bf 26}, 13 (2002)
  [hep-ph/0206068].
  
\bibitem{Hoferichter:2009ez} 
  M.~Hoferichter, B.~Kubis and U.-G.~Mei{\ss}ner,
  Phys.\ Lett.\ B {\bf 678}, 65 (2009)
  [arXiv:0903.3890 [hep-ph]].
  
\bibitem{Hoferichter:2009gn} 
  M.~Hoferichter, B.~Kubis and U.-G.~Mei{\ss}ner,
  Nucl.\ Phys.\ A {\bf 833}, 18 (2010)
  [arXiv:0909.4390 [hep-ph]].
  
\bibitem{Hoferichter:2012bz} 
  M.~Hoferichter {\it et al.},
  PoS {\bf CD  12}, 093 (2013)
  [arXiv:1211.1145 [nucl-th]].
  
\bibitem{PDG} 
  K.~A.~Olive {\it et al.}  [Particle Data Group Collaboration],
  Chin.\ Phys.\ C {\bf 38}, 090001 (2014).
  
\bibitem{Bernard:1995dp} 
  V.~Bernard, N.~Kaiser and U.-G.~Mei\ss ner,
  Int.\ J.\ Mod.\ Phys.\ E {\bf 4}, 193 (1995)
  [hep-ph/9501384].
  
\bibitem{Bernard:1996gq} 
  V.~Bernard, N.~Kaiser and U.-G.~Mei\ss ner,
  Nucl.\ Phys.\ A {\bf 615}, 483 (1997)
  [hep-ph/9611253].
  
\bibitem{Becher:1999he} 
  T.~Becher and H.~Leutwyler,
  Eur.\ Phys.\ J.\ C {\bf 9}, 643 (1999)
  [hep-ph/9901384].
  
\bibitem{Hemmert:1997ye} 
  T.~R.~Hemmert, B.~R.~Holstein and J.~Kambor,
  J.\ Phys.\ G {\bf 24}, 1831 (1998)
  [hep-ph/9712496].
  
\bibitem{Dashen:1993as} 
  R.~F.~Dashen and A.~V.~Manohar,
  Phys.\ Lett.\ B {\bf 315}, 425 (1993)
  [hep-ph/9307241].
  
\bibitem{Bernard:1995pa} 
  V.~Bernard, N.~Kaiser and U.-G.~Mei\ss ner,
  Phys.\ Rev.\ C {\bf 52}, 2185 (1995)
  [hep-ph/9506204].
  
\bibitem{Kaiser:2001pc} 
  N.~Kaiser,
  Phys.\ Rev.\ C {\bf 64}, 057001 (2001)
  [nucl-th/0107064].
  
\bibitem{Entem:2014msa} 
  D.~R.~Entem, N.~Kaiser, R.~Machleidt and Y.~Nosyk,
  Phys.\ Rev.\ C {\bf 91}, 014002 (2015)
  [arXiv:1411.5335 [nucl-th]].
  
\bibitem{Epelbaum:2014sza} 
  E.~Epelbaum, H.~Krebs and U.-G.~Mei{\ss}ner,
  Phys.\ Rev.\ Lett.\  {\bf 115}, 122301 (2015)
  [arXiv:1412.4623 [nucl-th]].
  
\bibitem{Epelbaum:2014efa} 
  E.~Epelbaum, H.~Krebs and U.-G.~Mei{\ss}ner,
  Eur.\ Phys.\ J.\ A {\bf 51}, 53 (2015)
  [arXiv:1412.0142 [nucl-th]].
  
\bibitem{Furnstahl:2015rha} 
  R.~J.~Furnstahl, N.~Klco, D.~R.~Phillips and S.~Wesolowski,
  Phys.\ Rev.\ C {\bf 92}, 024005 (2015)
  [arXiv:1506.01343 [nucl-th]].
  
\end{thebibliography}
\end{document}